\documentclass[3p]{elsarticle}
\usepackage{hyperref}
\hypersetup{colorlinks=true, urlcolor=blue}
\usepackage{cleveref}
\usepackage{epsfig}
\usepackage{newlfont}
\usepackage{amssymb}
\usepackage{amsfonts}
\usepackage{amsmath}
\usepackage{bm}
\usepackage{subfigure}

\def\lsim{\mathrel{\rlap{\lower4pt\hbox{\hskip1pt$\sim$}}
    \raise1pt\hbox{$<$}}}                % less than or approx. symbol
\def\gsim{\mathrel{\rlap{\lower4pt\hbox{\hskip1pt$\sim$}}
    \raise1pt\hbox{$>$}}}                % greater than or approx. symbol

\journal{Annals of Physics}

\begin{document}

\renewcommand*{\DefineNamedColor}[4]{%
   \textcolor[named]{#2}{\rule{7mm}{7mm}}\quad
  \texttt{#2}\strut\\}

\definecolor{red}{rgb}{1,0,0}
\definecolor{cyan}{cmyk}{1,0,0,0}

\title{Controllable quantum correlations of two-photon states generated using classically driven three-level atoms}

\author[sps]{Himadri Shekhar Dhar}

\author[iitr]{Subhashish Banerjee}

\author[sps]{Arpita Chatterjee}

\author[sps,snu]{Rupamanjari Ghosh\corref{cor1}}
\ead{rghosh.jnu@gmail.com}

\cortext[cor1]{Corresponding author}

\address[sps]{School of Physical Sciences, Jawaharlal Nehru University, New Delhi 110067, India}

\address[iitr]{Indian Institute of Technology Rajasthan, Jodhpur 342011, India}

\address[snu]{School of Natural Sciences, Shiv Nadar University, Gautam Budh Nagar, UP 203207, India}

\date{7 June 2012}

\begin{abstract}
We investigate the dynamics of two-photon correlations generated by the interaction of a three-level atom in the $\Xi$, $\Lambda$ or V configuration, with two classical external driving fields, under the rotating-wave approximation, in the presence of level decays. Using the example of a rubidium atom in each configuration, with field strengths validating the single-photon approximation, we compute measurement based correlations, such as measurement induced disturbance (MID), quantum discord (QD), and quantum work deficit (WD), and compare the results with that of quantum entanglement (concurrence). Certain correlation properties observed are generic, model independent and consistent with known results, e.g., MID is an upper bound on QD, QD and WD are monotonic, and the generic correlation behavior is strongly affected by the purity of the photon states. We observe that the qualitative hierarchy, monotonicity and steady-state behavior of the correlations can be controlled by the choice of parameters such as atomic decay constants and external driving field strengths. We point out how particular configurations are better suited at generating monotonic correlations in specific regimes and how the steady-state correlation behavior and hierarchy are affected by the population dynamics of the density matrix for different parameters. The possibility of using well studied quantum optical systems such as the three-level atom to generate, characterize and parametrically control mixed state quantum correlations establishes an important step in the direction of their implementation in quantum information tasks.
\end{abstract}


\maketitle

\section{Introduction}
\label{intro}

The interaction of atomic systems and external electromagnetic fields is a principal source for the generation and classification of quantum correlations \cite{Cohen}. The quantum nature of these atom-photon systems and the ability to implement such systems in controlled experimental settings make them important tools in the study of nonclassical features \cite{opt}. From the perspective of quantum information theory (QIT), atomic systems are the quintessential computational hardware needed for the future implementation of quantum information protocols \cite{atomqi, mem}, and photons are the basic building blocks of quantum communication \cite{photonqi} and cryptography \cite{crypt}. Hence, the generation and manipulation of nonclassical correlations in complex atomic systems interacting with radiation fields is one of the most challenging aspects of future applications of QIT.

The various popular indicators of nonclassicality and measures of quantum correlations are dependent on the theoretical perspectives invoked for their quantifications and are often not consistent when the interacting states have sufficient amount of mixedness \cite{Peters04}. The main dichotomy in the definition of quantum correlations arises from the question of what constitutes quantumness. The extensively studied entanglement-separability \cite{Hor} criterion to define quantum correlations stems from the understanding that the main feature of quantumness arises from the superposition principle \cite{sup}. Another feature to have received widespread attention in recent times is the definition of nonclassicality or quantumness on the basis of measurement based correlations. Such definitions take into account the fact that an important feature of quantumness in nature arises from noncommutativity of operators \cite{Luo}. Any physical measurement on a quantum system disturbs the noncommutative nature of the system and thus effectively erases quantum correlations. Information theoretic measures such as quantum discord (QD) \cite{QD, zur}, quantum work deficit (WD) \cite{QID} and measurement induced disturbance (MID) \cite{Luo} are based on the unique role of measurement in quantum physics.

The nonclassical properties of three-level atomic systems have been well studied in quantum optics for understanding quantum-coherence phenomena such as electromagnetically-induced transparency (EIT) \cite{EIT}, lasing without inversion \cite{LWI}, and coherent trapping \cite{scully97}. Three-level atoms interacting with low-strength driving fields, similar to EIT systems, have been used to generate entangled two-mode photon states which can be suitably manipulated to yield desired correlations \cite{SN}. The knowledge of nonclassical correlations carried by emitted photons in atomic systems may prove immensely useful in designing future QIT systems for communications and computation. The preparation and characterization of mixed quantum correlated states is a very challenging prospect with regards to generation of quantum states. Well known processes like parametric down-conversion \cite{PDC} generate pure entangled states with poor conversion efficiency and few control parameters. Quantum states generated by processes like resonance fluorescence have low signal-to-noise ratio with poor control over the emission statistics \cite{SN}. Quantum optical processes like coherent superposition operations have been used to generate and characterize nonclassicality and entangled states \cite{Non}. The use of generic quantum optical models, such as semiclassical three-level atomic systems, that can be experimentally implemented and observed, can serve as an important tool to generate, investigate, verify and control nonclassical correlations and their features.

\begin{figure}
%\label{config}
\begin{center}
\mbox{\subfigure[]{\includegraphics[scale=1]{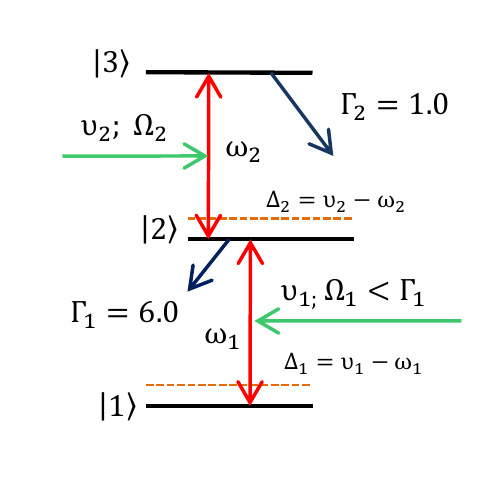}\label{fig1.1}}\qquad
      \subfigure[]{\includegraphics[scale=1]{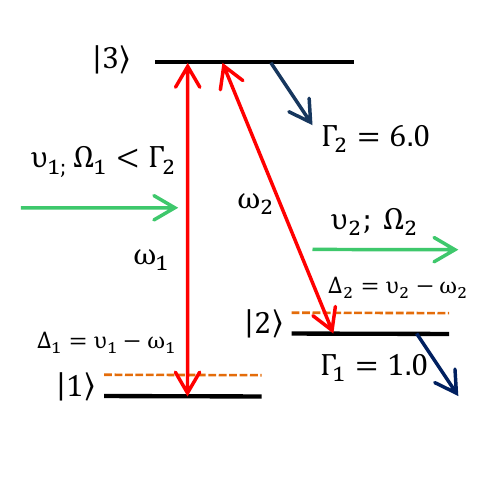}\label{fig1.2}}\qquad
       \subfigure[]{\includegraphics[scale=1]{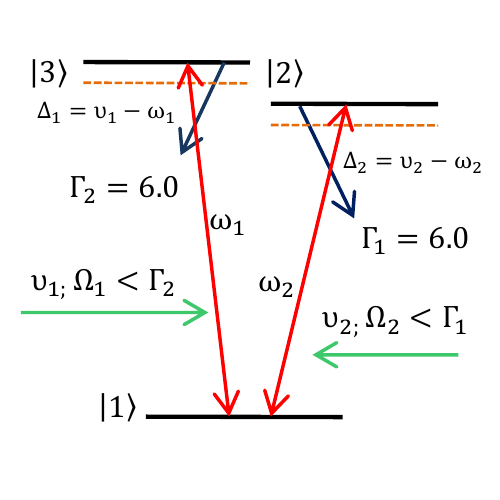}\label{fig1.3}}}
\caption{(Color online) A three-level atom in the (a) $\Xi$, (b) $\Lambda$, and (c) $V$ configuration. $\Gamma_1$ and $\Gamma_2$ are the decay constants (numbers shown in units of MHz) of the levels $|2\rangle$ and $|3\rangle$. $\nu_1$, $\nu_2$, and $\Omega_1$, $\Omega_2$ are the optical frequencies and the Rabi frequencies of the two near-resonant driving fields. $\omega_1$ and $\omega_2$ are the two atomic transition frequencies. $\Delta_1$ and $\Delta_2$ are the field detunings, set to zero in this work.}
\end{center}
\end{figure}

In this paper, we investigate the nonclassical correlation properties of the photon states emitted from a three-level atomic system interacting with two classical driving fields. The interactions generate two-mode single photon states, arising from two controlled coherent transitions connecting the three levels, under the single photon approximation (SPA) \cite{SPA}. The system can be set up in three different configurations, $\Xi$, $\Lambda$ and V. We exhaustively study the correlation properties of the emitted two-mode photons, and characterize the dynamics of the different measures of nonclassical correlations. We establish a qualitative relation between the two different theoretical classes of correlation measures, namely, entanglement and the measurement-based correlations such as MID, QD and WD. The control parameters in the system enable us to define specific regimes where certain correlations are enhanced based on the nature of the output photon states. We also analyze certain interesting features of the correlations generated by the interaction that throw light into the monotonicity and hierarchy of the set of measures used. The arrangement of the paper is as follows. We briefly discuss the different configurations of the three-level atom in Sec.\,\ref{sec2}. Then we have a short segment, in Sec.\,\ref{sec3}, defining the different correlation measures. In Sec.\,\ref{sec4}, we define the theoretical model used and the working approximations considered in the analysis. In Sec.\,\ref{sec5}, the numerical results obtained from the theoretical model are analyzed. We conclude in Sec.\,\ref{sec6}, with a summary of the results obtained and their possible ramifications.

\section{The three-level atom}
\label{sec2}
In this section, we briefly review our system. A three-level atom can be used in three different configurations, namely, $\Xi$, $\Lambda$ and V \cite{Eberly, level}. As a specific example, we focus on a gas of rubidium (Rb) atoms \cite{Ban}. The energy levels $5S_{1/2}$, $5P_{3/2}$ and $5D_{5/2}$ of Rb can be suitably used to generate each of the three configurations shown in Fig.\,\ref{fig1.1}-\subref{fig1.3}, as elaborated in the subsections below. Level $5S_{1/2}$ is the ground state and does not decay. Level $5D_{5/2}$ is metastable, with a decay rate of about 1.0 MHz.
%and we scale all rate/frequency values with the metastable level decay rate, which is about 1 MHz.
The decay of levels $5D_{5/2}$ and $5P_{3/2}$ are at
%(scaled)
rates of $\Gamma_D \approx$ 1.0 MHz and $\Gamma_P \approx$ 6.0 MHz, respectively \cite{Ban}.
The conditions on the driving field Rabi frequencies $\Omega_i$ shown in Fig.\,\ref{fig1.1}-\subref{fig1.3} are explained later in Sec.\,\ref{ssec4.3}.

\subsection{The $\Xi$ system}
\label{ssec2.1}
The cascade $\Xi$ system (Fig.\,\ref{fig1.1}) uses the allowed dipole transitions $|1\rangle \leftrightarrow |2\rangle$ and $|2\rangle \leftrightarrow |3\rangle$, with two classical fields of Rabi frequencies $\Omega_1$ and $\Omega_2$ driving these transitions, respectively. The field detunings are $\Delta_1$ and $\Delta_2$, set to zero throughout our analysis for near-resonant fields. The transition $|1\rangle \leftrightarrow |3\rangle$ is dipole forbidden. Levels $|1\rangle$, $|2\rangle$ and $|3\rangle$ correspond to the atomic levels $5S_{1/2}$, $5P_{3/2}$ and $5D_{5/2}$ of the Rb atom, respectively. Thus the decay rates of $|3\rangle$ and $|2\rangle$ are $\Gamma_2 \equiv \Gamma_D$ = 1.0 MHz and $\Gamma_1 \equiv \Gamma_P$ = 6.0 MHz, respectively. Level $P_{3/2}$ serves as the shared level $|2\rangle$ during the interaction. The initial atomic state is ground state ($|1\rangle$) populated and the levels $|2\rangle$ and $|3\rangle$ are unpopulated. $\Xi$ systems have been extensively used in coherent population trapping \cite{CPT} and also in experiments to achieve laser cooling in trapped ions \cite{cool}.

\subsection{The $\Lambda$ system}
\label{ssec2.2}
The $\Lambda$ system configuration can be obtained by folding the $\Xi$, with levels $5S_{1/2}$, $5P_{3/2}$ and $5D_{5/2}$ of the Rb atom now marked as levels $|1\rangle$, $|3\rangle$ and $|2\rangle$, respectively, as shown in Fig.\,\ref{fig1.2}. With this identification for Rb, we observe that level $|2\rangle$ is energetically higher than level $|3\rangle$. This corresponds to a negative transition frequency $\omega_2$. The rotating wave approximation (RWA) thus holds for the negative frequency term of the field in the Hamiltonian \cite{Ban}. Hence, the transition from level $|3\rangle$ to $|2\rangle$ annihilates a photon instead of creating a photon. The allowed dipole transitions are now $|1\rangle \leftrightarrow |3\rangle$ and $|2\rangle \leftrightarrow |3\rangle$, with two driving fields with Rabi frequencies $\Omega_1$ and $\Omega_2$ now acting on these transitions. Level $|3\rangle$ is the shared level, and the transition $|1\rangle \leftrightarrow |2\rangle$ is now dipole forbidden. The decay rates of $|2\rangle$ and $|3\rangle$ are $\Gamma_1 \equiv \Gamma_D$ = 1.0 MHz and $\Gamma_2 \equiv \Gamma_P$ = 6.0 MHz, respectively. The initial atomic system is again ground state ($|1\rangle$) populated, and the detunings are taken to be zero. The interactions of the three levels are distinctly different from the $\Xi$ system, and hence can be associated with different nonclassical behaviors. $\Lambda$ systems have been extensively used in demonstrating diverse coherent phenomena such as stimulated raman adiabatic passage \cite{STIRAP} and electromagnetically induced transparency (EIT) \cite{EIT}.

\subsection{The V system}
\label{ssec2.3}
The configuration of the V system (Fig.\,\ref{fig1.3}) is considerably different from the $\Xi$ and the $\Lambda$ systems. This is due to the fact that the shared level in the V system is the ground state. For the V system using Rb, we consider level $5S_{1/2}$ as the shared ground level ($|1\rangle$) and two hyperfine levels of $5P_{3/2}$ as the two-excited levels ($|2\rangle$ and $|3\rangle$). Hence, $\Gamma_1$ = $\Gamma_2 \equiv \Gamma_P$ = 6.0 MHz. The allowed transitions are $|1\rangle \leftrightarrow |3\rangle$ and $|1\rangle \leftrightarrow |2\rangle$, driven by the classical fields of Rabi frequencies $\Omega_1$ and $\Omega_2$, respectively. The transition $|2\rangle \leftrightarrow |3\rangle$ is dipole-forbidden, i.e., the ground state excitations take the system to two excited levels that cannot be coupled, and interactions are thus limited to ground state transitions. The initial atomic system is again ground state ($|1\rangle$) populated, and the detunings are set to zero. V systems are widely used to study nonclassical phenomena such as quantum jumps \cite{jump}, quantum Zeno effect \cite{zeno} and quantum beats \cite{scully97}.

\section{Quantum correlations}
\label{sec3}
We compare measures of quantum correlations that are defined from two different perspectives. Information-theoretic measures, such as MID \cite{Luo}, QD \cite{QD,zur} and WD \cite{QID}, are based on the modification of quantum correlations upon measurement. These correlations calculate the difference in some specific property between quantum states and their measured classical projections to give us a measure of the nonclassicality. On the other hand, there are nonclassical measures based on the entanglement-separability criterion. We use concurrence \cite{con} as an example of entanglement monotone \cite{Hor}.

\subsection{Measurement based correlations}
\label{ssec3.1}
\emph{Measurement Induced Disturbance (MID)}: It is derived from the understanding that a truly classical state, with respect to some measurement, will remain unchanged after the measurement \cite{Luo}. Let us consider a bipartite density matrix $\rho_{ab}$. If {$B_i^a$} and {$B_j^b$} are complete von Neumann measurements (one dimensional projections) for subsystems $a$ and $b$, respectively, for a classical state,
\begin{eqnarray}
\label{MID}
\rho_{ab}=\sum_{ij} B_i^a \otimes B_j^b ~ \rho_{ab} ~ B_i^a \otimes B_j^b .
\end{eqnarray}
The states $\rho_{ab}$ that do not satisfy (\ref{MID}) are essentially quantum in nature. MID measures the quantumness in a bipartite state $\rho_{ab}$ by measuring the difference in the quantum mutual information between the state $\rho_{ab}$ and its least disturbed classical state obtained by the measurement, $\rho_{class}=\sum_{i} \mathcal{B}_i ~\rho_{ab}~ \mathcal{B}_i$, where {$\mathcal{B}_i$} are the spectral projections of the state $\rho_{ab}$. Thus
\begin{equation}
\mathcal{M}(\rho_{ab})= \mathcal{I}(\rho_{ab})-\mathcal{I}(\rho_{class}) ,
\end{equation}
where $\mathcal{I}(\rho_{ab}) \equiv S(\rho_a)+ S(\rho_b)- S(\rho_{ab})$ is the quantum mutual information \cite{mut}. $S(\rho)= - \mbox{tr} (\rho \log_2~ \rho)$ is the von Neumann entropy of a quantum state $\rho$. $\rho_a$ and $\rho_b$ are the reduced density matrices of the subsystem $a$ and $b$, respectively. $\mathcal{M}(\rho_{ab})$ is the nonclassical measure of MID \cite{Luo}. Unlike other measures of correlation based on projective measurements, MID does not introduce any optimization on the measured states. Hence, MID serves as an upper bound on other measurement based correlations \cite{subh}.

\emph{Quantum Discord (QD)}: It is defined as the difference between two classically equivalent expressions for mutual information when extended to the quantum regime \cite{QD, zur}. For the density operator $\rho_{ab}$, the expressions for quantum mutual information are
\begin{eqnarray}
\label{mutual}
I(\rho_{ab})&\equiv& S(\rho_a)+ S(\rho_b)- S(\rho_{ab}) \nonumber\\
 & \neq & J(\rho_{ab}) \equiv S(\rho_a) - S(\rho_{a|b}),
\end{eqnarray}
where $S(\rho_{ab})$, as defined earlier, is the von Neumann entropy. $I(\rho_{ab})$ is the quantum mutual information \cite{mut} and $S(\rho_{a|b})$ is the quantum conditional entropy \cite{cond}. The two expressions in (\ref{mutual}) are equal in the classical regime.

To calculate the quantum conditional entropy, $S(\rho_{a|b})$, we use the one-qubit orthonormal projection basis: $\left|i_1\right\rangle = \cos \frac{\theta}{2} \left|0\right\rangle + e^{i\phi} \sin \frac{\theta}{2} \left|1\right\rangle$ and $\left|i_2\right\rangle =  e^{-i\phi} \sin \frac{\theta}{2} \left|0\right\rangle + \cos \frac{\theta}{2} \left|1\right\rangle$, with $\langle i|j \rangle = \delta_{ij}$, $i, j = 0, 1$. The projection density operators are given by the relations $B_1 = \left|i_1\rangle\langle i_1\right|$ and $B_2 = \left|i_2\rangle\langle i_2\right|$. On measuring the subsystem $b$ using the above set of projection operators, the post-measurement states are $\rho^i_{ab}= \frac{1}{p_i}(\mathbb{I}_a \otimes\,\, B_i ~\rho_{ab}~\, \mathbb{I}_a \otimes B_i)$, where $p_i = \mbox{tr}_{ab}(\mathbb{I}_a \otimes B_i \,\,\rho_{ab}~\, \mathbb{I}_a \otimes B_i)$, and $\mathbb{I}_a$ is the identity operator acting on the subsystem $a$. The quantum conditional entropy is then given by
\begin{equation}
\label{cond}
S(\rho_{a|b}) = \min_{\{B_i\}} \sum_i p_i S(\rho^i_{ab}) .
\end{equation}
QD can then be defined using relation (\ref{mutual}) and (\ref{cond}).
\begin{eqnarray}
J(\rho_{ab}) & = & S(\rho_a) - \min_{\{B_i\}} \sum_i p_i S(\rho^i_{ab}),\\
QD(\rho_{ab})& = & I(\rho_{ab}) - J(\rho_{ab}).
\end{eqnarray}

\emph{Quantum work deficit (WD)}: This concept is based on the fact that information is a thermodynamic resource \cite{QID}. It is defined as the amount of work (in terms of pure states) that can be extracted from a quantum bipartite system under a closed global operation and the amount that can be extracted using closed local operations and classical communications (CLOCC) \cite{QID2}. WD is the deficit in the two operations due to loss of nonclassical correlations while performing CLOCC. Hence, WD is a measure of nonclassical correlations.

Using the projection basis and the expressions for entropy defined previously, we can obtain the expressions for the global and CLOCC operations. Under the class of global operations, the amount of work extractable (in terms of pure states) is given by $I_G(\rho_{ab})= \text{log}_2 \text{dim} \cal{H} - S(\rho_{ab})$, where dim$\cal{H}$ is the dimension of the Hilbert space. The amount of work that can be extracted using CLOCC is dependent upon local unitary operations, local dephasing and classical communication of the dephased state. $\rho_{ab} \rightarrow  \sum_i B_i ~\rho_{ab} ~B_i$.
\begin{eqnarray}
\rho_{ab}^i &=& \sum_i(\mathbb{I}_a \otimes\,\, B_i ~\rho_{ab}~\, \mathbb{I}_a \otimes B_i) .
\end{eqnarray}
The amount of work that can be extracted using CLOCC is given by $I_L(\rho_{ab})= \text{log}_2 \text{dim} \cal{H} - \inf_{CLOCC}[S(\rho_{ab}^i)]$. WD is then defined by the following expression:
\begin{equation}
\Delta(\rho_{ab})=I_G(\rho_{ab})-I_L(\rho_{ab}) .
\end{equation}

\subsection{Entanglement measure}
\label{ssec3.2}
\emph{Concurrence}: Entanglement in a bipartite system can be measured using this entanglement monotone defined for mixed states of two qubits \cite{con}. Concurrence can be defined for a two-qubit density matrix $\rho(t)$ as
$C (\rho) = \mathrm{max}[0, \lambda_1-\lambda_2-\lambda_3-\lambda_4]$,
where $\lambda_i (i=1,2,3,4)$ are the square roots of the eigenvalues of the \textit{spin-flip} operator, $R=\rho(t) \tilde{\rho}(t)$, with
$\tilde{\rho}(t) = (\sigma_y \otimes \sigma_y)\rho(t) (\sigma_y \otimes \sigma_y)$, and $\sigma_y$ is the Pauli spin matrix,
\begin{eqnarray}
\sigma_y&=& \left(
             \begin{array}{cc}\nonumber
                  0 & -i\\
                   i & 0
             \end{array}
           \right).
\end{eqnarray}
Concurrence is closely related to the entanglement of formation. The entanglement of formation is a monotonically  increasing function of concurrence. Concurrence, however, is not a resource-based measure such as entanglement of formation \cite{woot}.

\section{Model}
\label{sec4}
Three-level atomic systems have been extensively used to study quantum and nonlinear features of the semiclassical atom-field system \cite{Eberly} and also the nonclassical nature of emitted radiation in such systems \cite{Clauser}. Entanglement properties of three-level atomic systems have also been investigated \cite{ent3level}. However, a general classification of information-theoretic correlations for such systems does not exist, and we wish to investigate and compare the features of the above quantum correlation features using this versatile system.

\subsection{The Hamiltonian}
\label{ssec4.1}
The Hamiltonian for a general three-level atom interacting with two classical driving fields, in the RWA, can be written as \cite{scully97}:
\begin{eqnarray}
\mathcal{H} &=& \mathcal{H}_0 +\mathcal{H}_I, \\\nonumber
\mathcal{H}_0 &=&  \hbar \omega_{11} |1\rangle\langle 1| + \hbar \omega_{22} |2\rangle\langle 2|+ \hbar \omega_{33} |3\rangle\langle 3|,  \\
\mathcal{H}_I &=& -\hbar /2~(\Omega_1 e^{-i\phi_1} e^{-i \nu_1 t} |m\rangle\langle n| + \Omega_2 e^{-i\phi_2} e^{-i \nu_2 t} |l\rangle\langle k| + \mbox{H.c.}),
\end{eqnarray}
where $\hbar\omega_{ii}$ is the energy of level $|i\rangle$ $(i=1,2,3)$; $\Omega_j e^{-i\phi_j}$ $(j=1,2)$ is the complex Rabi frequency corresponding to the classical driving field of frequency $\nu_j$, $m, n, l, k = \{1, 2, 3\}$ denote the three atomic levels as appropriate for the $\Xi$, $\Lambda$ or V configuration. For the $\Xi$ configuration, $(m, n, l, k)=(2, 1, 3, 2)$ correspond to the atomic transitions $|1\rangle \leftrightarrow |2\rangle$ and $|2\rangle \leftrightarrow |3\rangle$. For the $\Lambda$ configuration, $(m, n, l, k) = (3, 1, 3, 2)$ correspond to the transitions $|1\rangle \leftrightarrow |3\rangle$ and $|2\rangle \leftrightarrow |3\rangle$, and for the V configuration, $(m, n, l, k)=(3, 1, 2, 1)$ correspond to the transitions $|1\rangle \leftrightarrow |3\rangle$ and $|1\rangle \leftrightarrow |2\rangle$ (Fig.\,\ref{fig1.1}-\subref{fig1.3}).

\subsection{The Atomic Density Matrix}
\label{ssec4.2}
The state of the atomic system, at any time $t$, can be written in the following form:
\begin{equation}
\label{w}
|\psi(t)\rangle_A = C_1(t) e^{-i(\omega_{11}+\xi_1)t}|1\rangle + C_2(t) e^{-i(\omega_{22}+\xi_2)t}|2\rangle + C_3(t) e^{-i(\omega_{33}+\xi_3)t}|3\rangle ,
\end{equation}
where $\xi_i (i=1,2,3)$ are phases that depend on the detunings in a specific configuration. The detunings are defined as
\begin{eqnarray}
\Delta_1 &=& \nu_1 - \omega_1, \nonumber \\
\Delta_2 &=& \nu_2 - \omega_2, \nonumber
\end{eqnarray}
where $\omega_1= \omega_{mm}-\omega_{nn}$, and $\omega_2=\omega_{ll}-\omega_{kk}$, are the transition frequencies. ($m, n, l, k$) have been defined earlier and are different for the three configurations.

Using the wavefunction (\ref{w}), one can create a pure state atomic density matrix $\rho_A(t)$ which depends on the classical driving field frequencies $\nu_j (j=1,2)$.
%The evolution of the atomic state is also followed by the decay of the energy levels.
We take phenomenological parameters to denote spontaneous decays of the excited atomic levels \cite{scully97}. The parametrized decay terms may lead to spontaneously generated coherences in the decay paths \cite{SGC}. Such phenomenological decay models are commonly used in the study of quantum features in EIT \cite{Ban}, in quantum state tomography of emitted field states \cite{SN}, and in earlier studies on atom-photon entanglement \cite{entropy}. With the level decay terms, the dynamics of the system is in general mixed, and can be obtained using the von Neumann (quantum Liouville) equation of motion,
\begin{equation}
\label{VN}
\dot{\rho}_A(t)= -\frac{i}{\hbar}[\mathcal{H},\rho_A(t)]-\frac{1}{2}\{\Gamma,\rho_A(t)\},
\end{equation}
where the elements of the relaxation matrix $\Gamma$ are the decay rates, $\langle i|\Gamma |j \rangle = \Gamma_{i-1} \delta_{ij}$, $i, j = 1, 2, 3$. The time-evolved mixed atomic density matrix can be obtained provided the initial states of the atom (before interaction) are known.

\subsection{The Photon Density Matrix}
\label{ssec4.3}
The nonclassical nature of the emitted radiation is dependent on the interaction between the three-level atomic system and the two-mode classical driving fields. The desired output, in our case, is to limit the generation to single photons for the two modes emitted after the interaction, so that at any given time within the lifetime of the atom, there will exist two photon states for each mode. Thus the two-photon density matrix can be written as
\[ \rho_{ph} = \sum \rho_{ij,i'j'} |i j\rangle \langle i' j'|, \]
where $|i,j\rangle$ ($|i',j'\rangle$) stands for the two-photon states, with $i$ and $j$ ($i'$ and $j'$) = 0, 1 representing the number of photons in the first and second modes, respectively. Such a two-photon state can be achieved using the single photon approximation (SPA) \cite{SPA} within the rotating wave approximation (RWA) \cite{scully97}. The RWA ensures that a photon is created only when an atomic de-excitation takes place. The SPA is applied by ensuring that the excitation time (due to the driving field strength) is larger than the decay time. If $\Omega$ is the driving field Rabi frequency and $\Gamma$ is the atomic decay rate, we require $1/\Omega > 1/\Gamma$ for the SPA to be valid, so that the time taken for an atom to excite is much greater than the decay time and for small times only a single de-excitation will occur generating a single photon.\footnote{We consider evolution times much smaller than the lifetime of the atom, since the condition of single photon emission may not remain valid due to spontaneous decay over a longer period.} Hence, the ground state excitation strength $\Omega$ should be smaller than the upper-level decay rate $\Gamma$.
%\textcolor{red}{We consider evolution times much lower than the atomic lifetime of the atom such that the system never reaches a steady state where the single photon emission may no longer be valid due to spontaneous decay.}

The output state is thus a two-qubit (bipartite) photon state. For a semiclassical interaction involving atoms and driving fields, it has been shown that the density matrix of the output radiation state can be completely derived from the atomic density matrix \cite{SN}. Under the far-field approximation \cite{scully97, Man}, for an atom located at $\vec{r}_0$, the field operators of the emitted radiation at the point of detection $\vec{r}$ are proportional to the atomic spin operators at the retarded time $(t - |\vec{r} - \vec{r}_0|/c)$. This equivalence leads to an expression for the photon density matrix $\rho_{ph}(t)$ that is identical to the atomic density matrix at an earlier time, $\rho_A(t - r/c)$, calculated using the von Neumann equation of motion. The photon density matrix has a reduced rank three \cite{comm1}. It has been shown \cite{SN} that such an equivalence leads to the complete determination of the output photon state using quantum state tomography, where measurements can be made on either the atomic or the photonic operators. Since the photon states can be completely determined by the atomic density matrix, the coherence of the photon correlations is closely related to the evolution of the atomic state.

\section{Results and Analysis}
\label{sec5}

As mentioned earlier, the system we consider is a gas of Rb atoms. The three levels $5S_{1/2}$, $5P_{3/2}$ and $5D_{5/2}$ of the Rb atom are appropriated to obtain the $\Xi$, $\Lambda$ and V configurations. All rates and frequencies are rendered dimensionless by scaling with the metastable level decay rate ($\approx$ 1 MHz). The scaled decay rates of $5P_{3/2}$ and $5D_{5/2}$ are $\Gamma_P$ = 6.0 and $\Gamma_D$ = 1.0, respectively, and $5S_{1/2}$ is the ground state ($\Gamma_S$ = 0) \cite{Ban}. For desired results of two-mode single photon generation, we restrict ourselves to regimes that satisfy the SPA. In the $\Xi$ and $\Lambda$ configuration (Fig.\,\ref{fig1.1} and \subref{fig1.2}), since the shared level is $5P_{3/2}$, we take the ground state excitation field ($\Omega_1$) to be always less than the decay constant of $5P_{3/2}$ (in $\Xi$, $\Omega_1 < \Gamma_1 \equiv \Gamma_P$ = 6.0; in $\Lambda$, $\Omega_1 < \Gamma_2 \equiv \Gamma_P$ = 6.0). For the V configuration (Fig.\,\ref{fig1.3}), the ground state excitation leads to transitions to the hyperfine levels of $5P_{3/2}$, and hence both the driving fields $\Omega_1$ and $\Omega_2$ are less than the decay constant of $5P_{3/2}$ ($\Omega_{1,2} < \Gamma_{1,2} \equiv \Gamma_P$ = 6.0). In our analysis, we set the atom-field detunings to zero \cite{comm2}.
We calculate the measurement-based quantum correlations and entanglement of the output photon density matrix that can be derived using the atomic density matrix (Sec.\,\ref{ssec4.1}) obtained from the von Neumann equation of motion (\ref{VN}), in the three configurations.
%We present only specific regimes to highlight interesting features of the correlations in the interaction dynamics.
%%%\begin{figure}[h]
%%%%\label{fig:3}
%%%\begin{center}
%%%\subfigure[]{\includegraphics[width=6.2cm]{timeoutput1.pdf}\label{fig3.1}}\qquad\quad\quad\qquad
%%%\subfigure[]{\includegraphics[width=6.2cm]{timeoutput2.pdf}\label{fig3.2}}
%%%\caption{(Color online) Fixed time ($t$ = 1.0) MID (red continuous), discord (blue circles), work deficit (green squares), and concurrence (black dashed) in the $\Xi$ configuration as a function of the driving field strength $\Omega_2$. The field detunings are $\Delta_1$ = $\Delta_2$ = 0, and the phases of the Rabi frequencies are $\phi_1$ = $\phi_2$ = 0. The level decay rates are $\Gamma_1$ = 6.0, $\Gamma_2$ = 1.0. SPA for this configuration requires that $\Omega_1 < \Gamma_1$. One driving field strength $\Omega_1$ is fixed at (a) 1.5, and (b) 3.5. The inset shows the variation of population elements of the two-photon density matrix and its purity.}
%%%\label{fig3}
%%%\end{center}
%%%\end{figure}

\begin{figure}[htbp]
%\label{fig:2}
\begin{center}
\subfigure[]{\includegraphics[width=6.0cm]{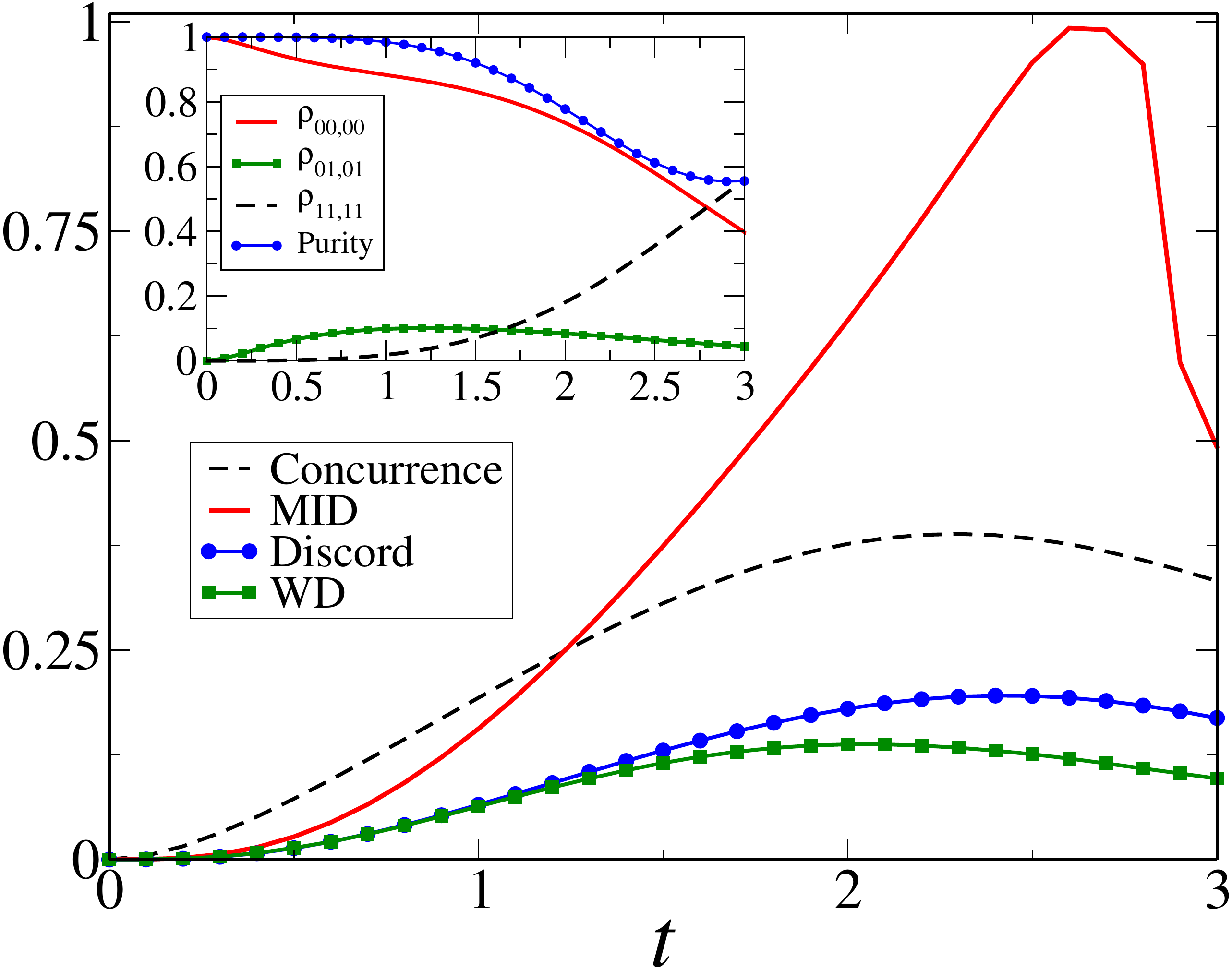}\label{fig2.1}}\qquad\quad\quad\qquad
      \subfigure[]{\includegraphics[width=6.0cm]{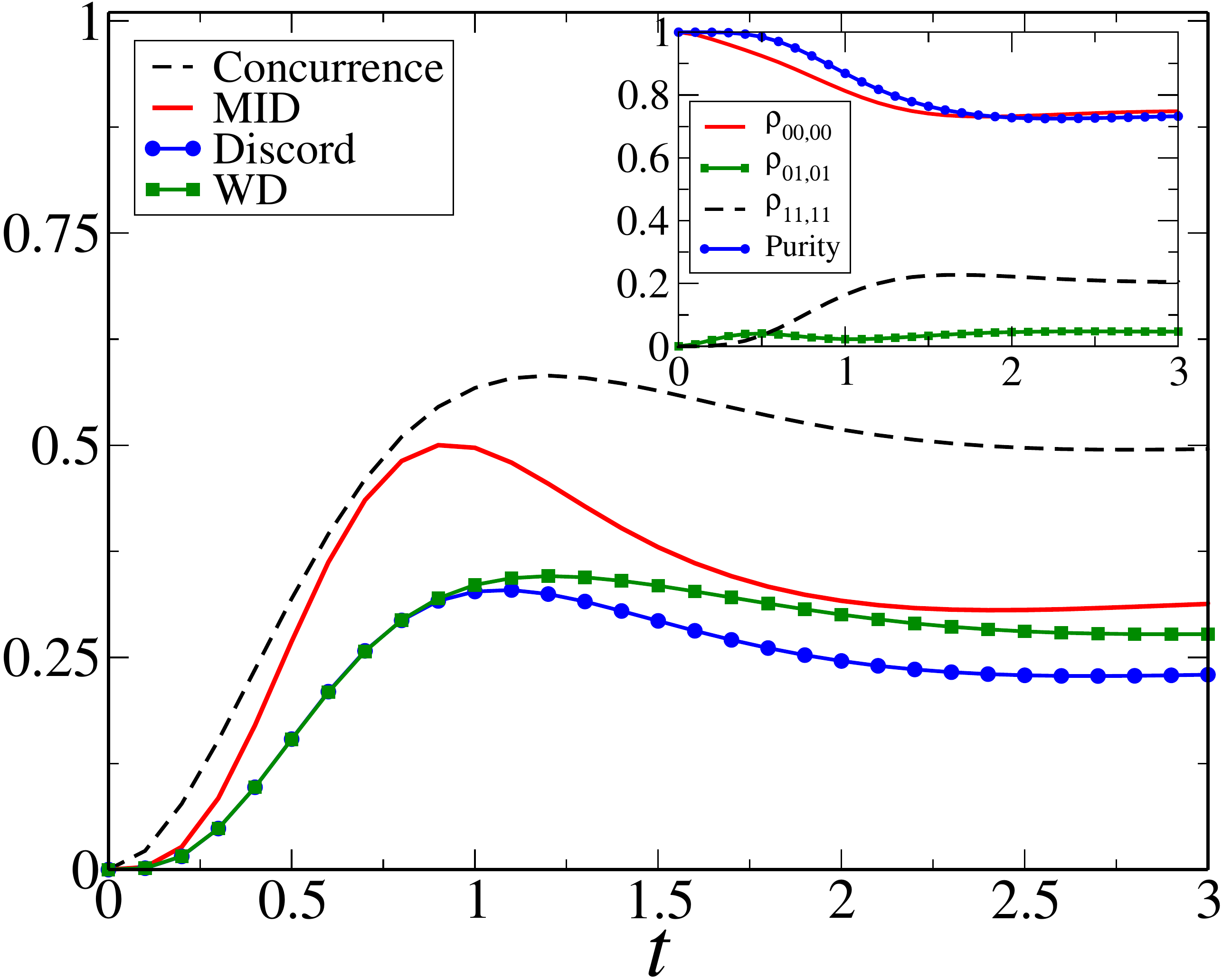}\label{fig2.2}}
\caption{(Color online) The time evolution for correlation measures MID (red continuous), discord (blue circles) and work deficit (green squares) along with the entanglement measure concurrence (black dashed) for the cascade ($\Xi$) configuration. The field detunings are $\Delta_1$ = $\Delta_2$ = 0, and the phases of the Rabi frequencies are $\phi_1$ = $\phi_2$ = 0. The level decay rates are $\Gamma_1$ = 6.0, $\Gamma_2$ = 1.0. SPA for this configuration requires that $\Omega_1 < \Gamma_1$. The driving field strengths are (a) $\Omega_1$ = 2.0, $\Omega_2$ = 1.0, and (b) $\Omega_1$ = 2.0, $\Omega_2$ = 5.0. The inset shows the evolution of population elements of the two-photon density matrix and its purity.}
\label{fig2}
\end{center}
\end{figure}

\begin{figure}[h]
%\label{fig:3}
\begin{center}
\subfigure[]{\includegraphics[width=6.2cm]{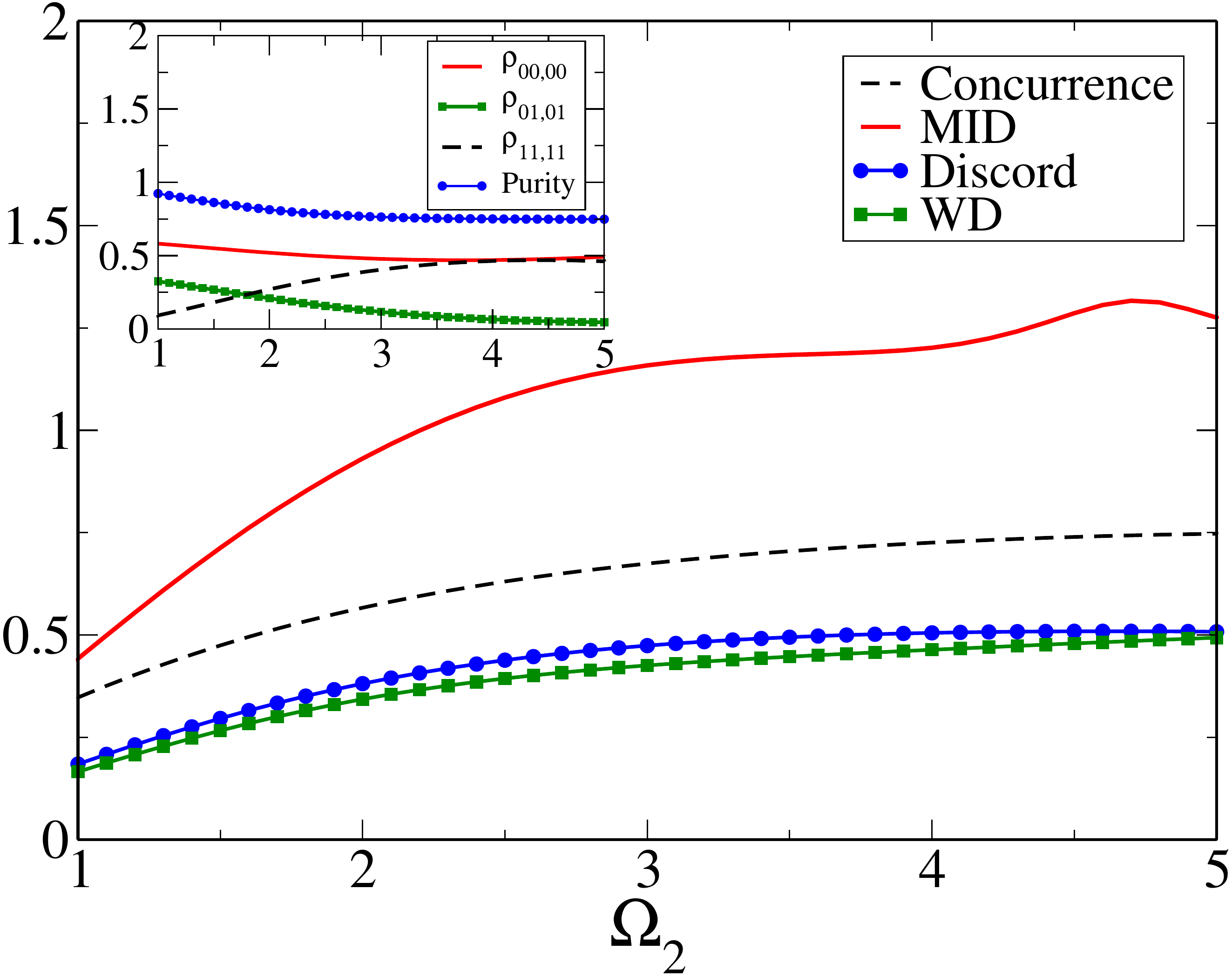}\label{fig3.1}}\qquad\quad\quad\qquad
\subfigure[]{\includegraphics[width=6.2cm]{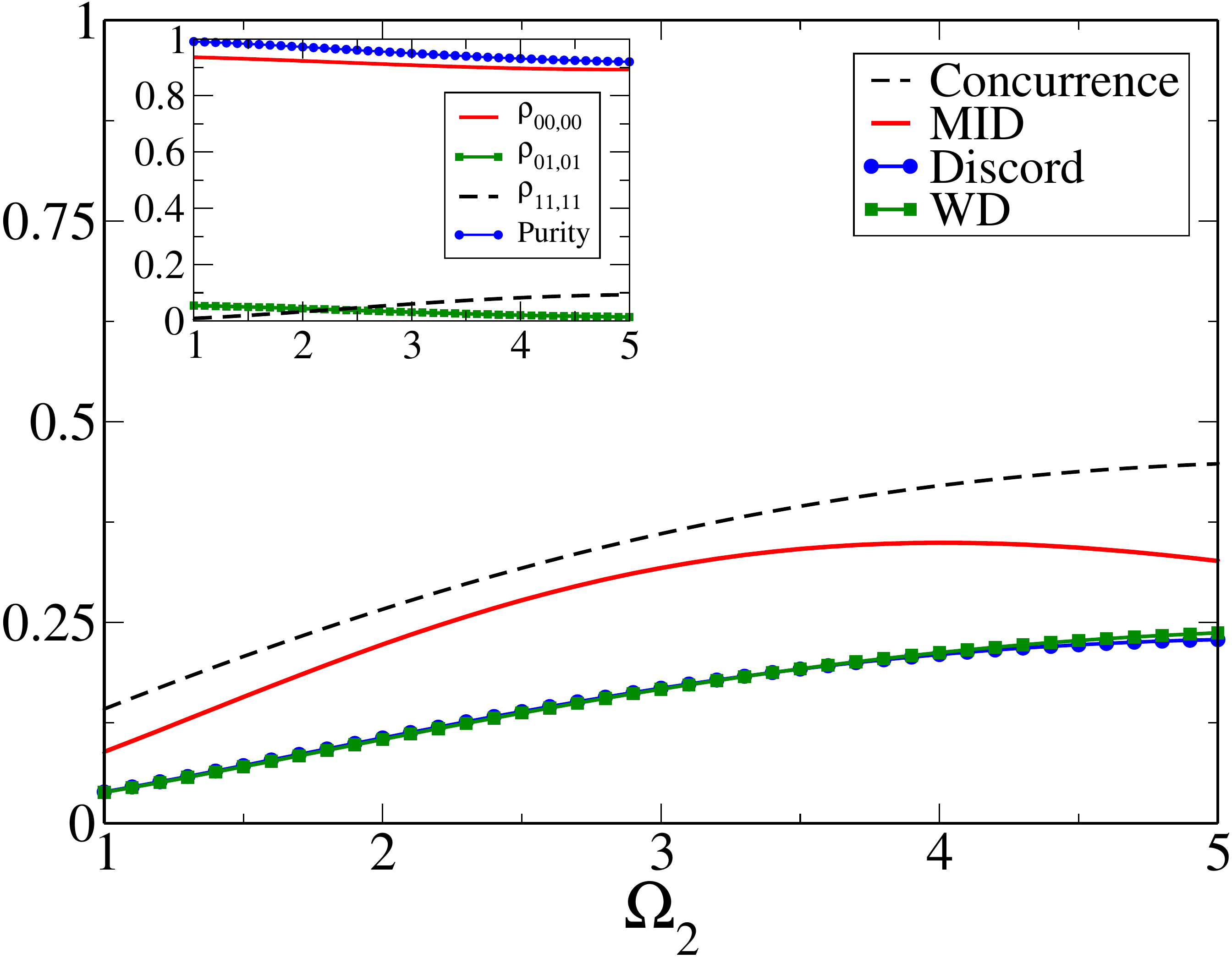}\label{fig3.2}}
\caption{(Color online) Fixed time ($t$ = 1.0) MID (red continuous), discord (blue circles), work deficit (green squares), and concurrence (black dashed) in the $\Xi$ configuration as a function of the driving field strength $\Omega_2$. The field detunings are $\Delta_1$ = $\Delta_2$ = 0, and the phases of the Rabi frequencies are $\phi_1$ = $\phi_2$ = 0. The level decay rates are $\Gamma_1$ = 6.0, $\Gamma_2$ = 1.0. SPA for this configuration requires that $\Omega_1 < \Gamma_1$. One driving field strength $\Omega_1$ is fixed at (a) 1.5, and (b) 3.5. The inset shows the variation of population elements of the two-photon density matrix and its purity.}
\label{fig3}
\end{center}
\end{figure}
We discuss the interesting quantum correlations generated in the two-mode photon state in specific parameter regimes using the $\Xi$ configuration as the reference. Some general observations can be made that are consistent with known results: MID always serves as an upper bound on the other measurement based correlations, such as, QD and WD \cite{subh}. The evolution of MID with respect to concurrence can be varied using the control parameters. There are two specific parameter regimes that correspond to two different hierarchies in the photon correlation. In Fig.\,\ref{fig2}, we consider the $\Xi$ configuration in two specific regimes of the driving classical fields. The decay constants for the $\Xi$ configuration are $\Gamma_1$ = 6.0, $\Gamma_2$ = 1.0. Hence, the driving field strengths are in the range ($\Omega_1, \Omega_2) <$ 6.0. The detunings and the Rabi frequency phases have been set to zero. In Fig.\,\ref{fig2.1}, we consider the regime where $\Omega_1 > \Omega_2$. Fig.\,\ref{fig2.2}, corresponds to the field regime $\Omega_1 < \Omega_2$ ($\Omega_1 = 2.0, \Omega_2 = 5.0$). For $\Omega_1 > \Omega_2$, MID forms a non-monotonic upper bound on concurrence at times $t >$ 1.0. For the field regime $\Omega_1 < \Omega_2$ ($\Omega_1 = 2.0, \Omega_2 = 5.0$), concurrence forms a monotonic upper bound on the measurement-based correlations. We observe that the behavior of the correlations is closely related to the dynamics of the populations (inset of Fig.\,\ref{fig2}). The non-monotonic behavior of MID is associated with the population difference in the two photon modes $|00\rangle$ and $|11\rangle$. It is clear from the plots that the sudden increase in MID occurs when the populations of the modes $|00\rangle$ and $|11\rangle$ are nearly equal. This could be due to the fact that the non-optimization of the correlation measure in MID is skewed in these regions. For cases where MID is monotonic with the other measurement based measures, the population is distinctly unequal. Observing the purity in these regimes, one can state that the monotonicity is observed at higher levels of purity.

\begin{figure}[htbp]
%\label{fig:2}
\begin{center}
\mbox{\subfigure[]{\includegraphics[width=6.2cm]{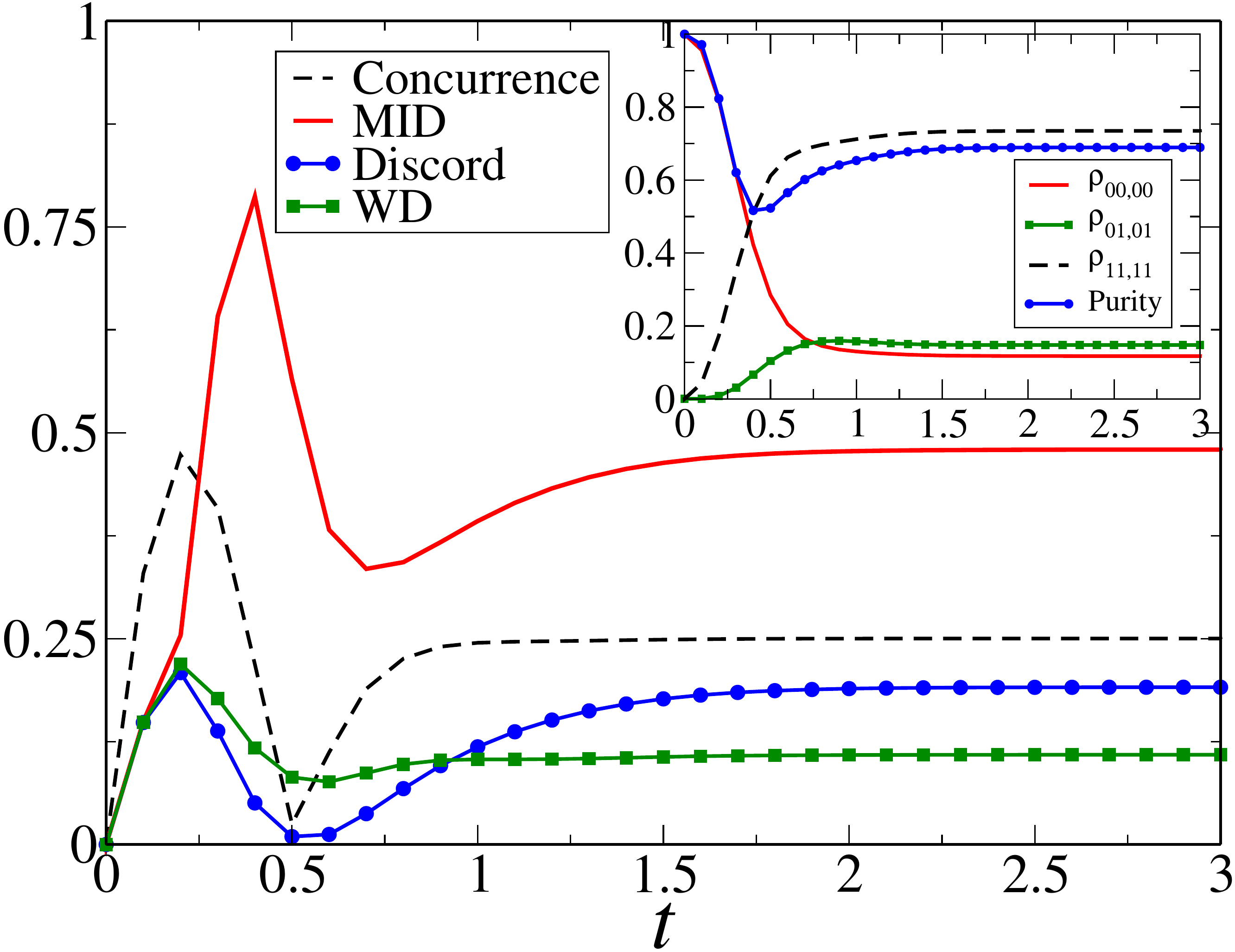}\label{fig4.1}}\qquad\quad\quad\qquad
      \subfigure[]{\includegraphics[width=6.2cm]{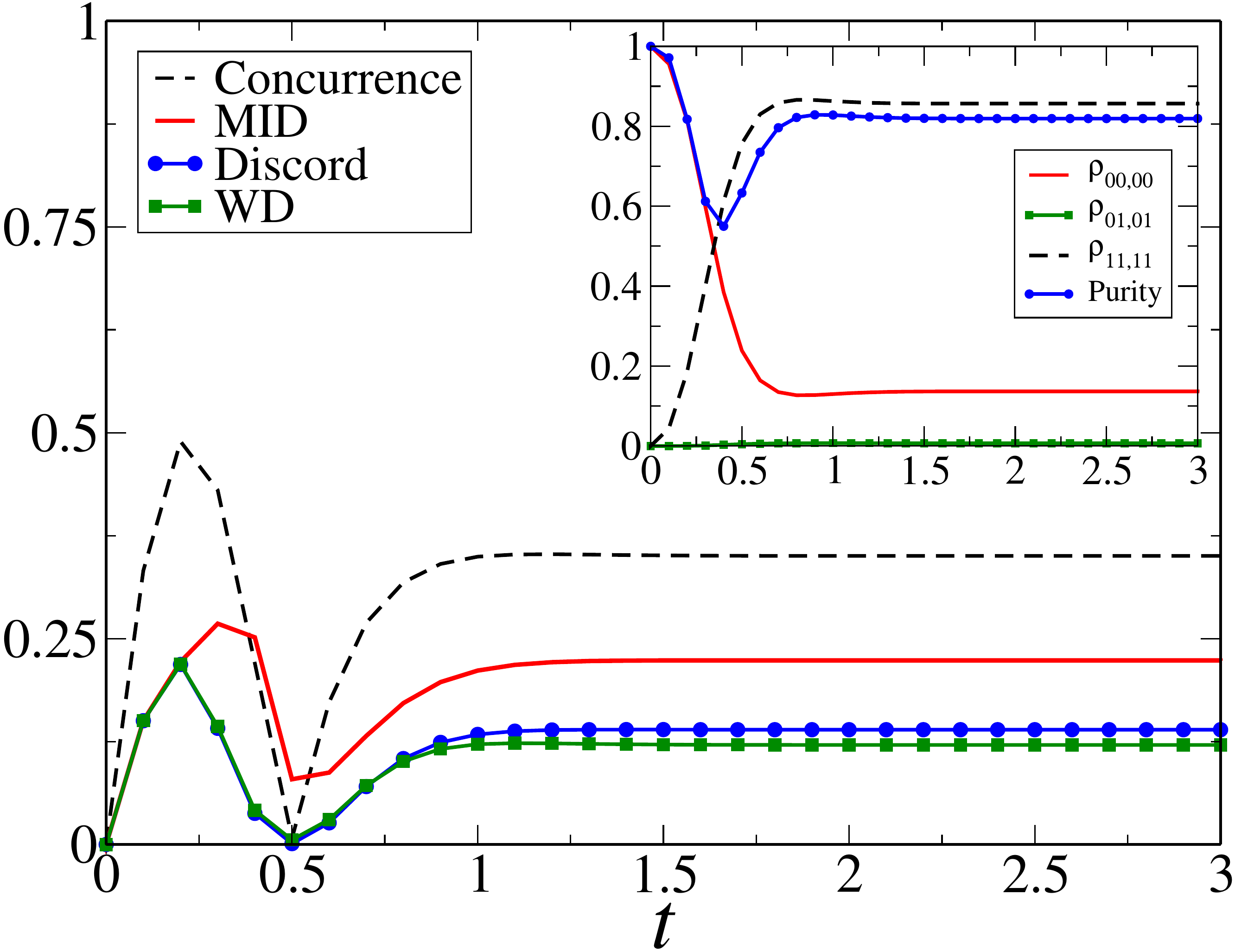}\label{fig4.2}}}
\caption{(Color online) The time evolution for MID (red continuous), discord (blue circles), work deficit (green squares) and concurrence (black dashed) for the $\Lambda$ configuration. The field detunings are $\Delta_1$ = $\Delta_2$ = 0, and the phases of the Rabi frequencies are $\phi_1$ = $\phi_2$ = 0. The level decay rates are $\Gamma_1$ = 1.0, $\Gamma_2$ = 6.0. SPA for this configuration requires that $\Omega_1 < \Gamma_2$. The driving field strengths are thus taken as (a) $\Omega_1$ = 4.0, $\Omega_2$ = 5.0, and (b) $\Omega_1$ = 4.0, $\Omega_2$ = 1.0. The inset shows the evolution of population elements of the two-photon density matrix and its purity.}
\label{fig4}
\end{center}
\end{figure}

A similar dichotomy in behavior can also be observed for fixed time dynamics of the system if the interaction is allowed to vary across driving field strengths. In Fig.\,\ref{fig3}, keeping the evolution time fixed ($t$ = 1.0) and varying the two classical driving field strengths, a similar behavior of the correlations is observed. MID is greater than concurrence and non-monotonic at times where the population levels are equal with significantly lower purity (Fig.\,\ref{fig3.1}) as compared to regimes with unequal populations and higher purity where the measurement based correlations are monotonic to concurrence (Fig.\,\ref{fig3.2}). Hence, we observe that the fixed time dynamics allows us to manipulate the correlation hierarchy by changing the ground-state driving field strength, $\Omega_1$. The generation of monotonic correlations can be controlled by using parameter regions that allow higher purity in the output photon state.

\begin{figure}[htbp]
%\label{fig:5}
\begin{center}
\includegraphics[width=6.2cm]{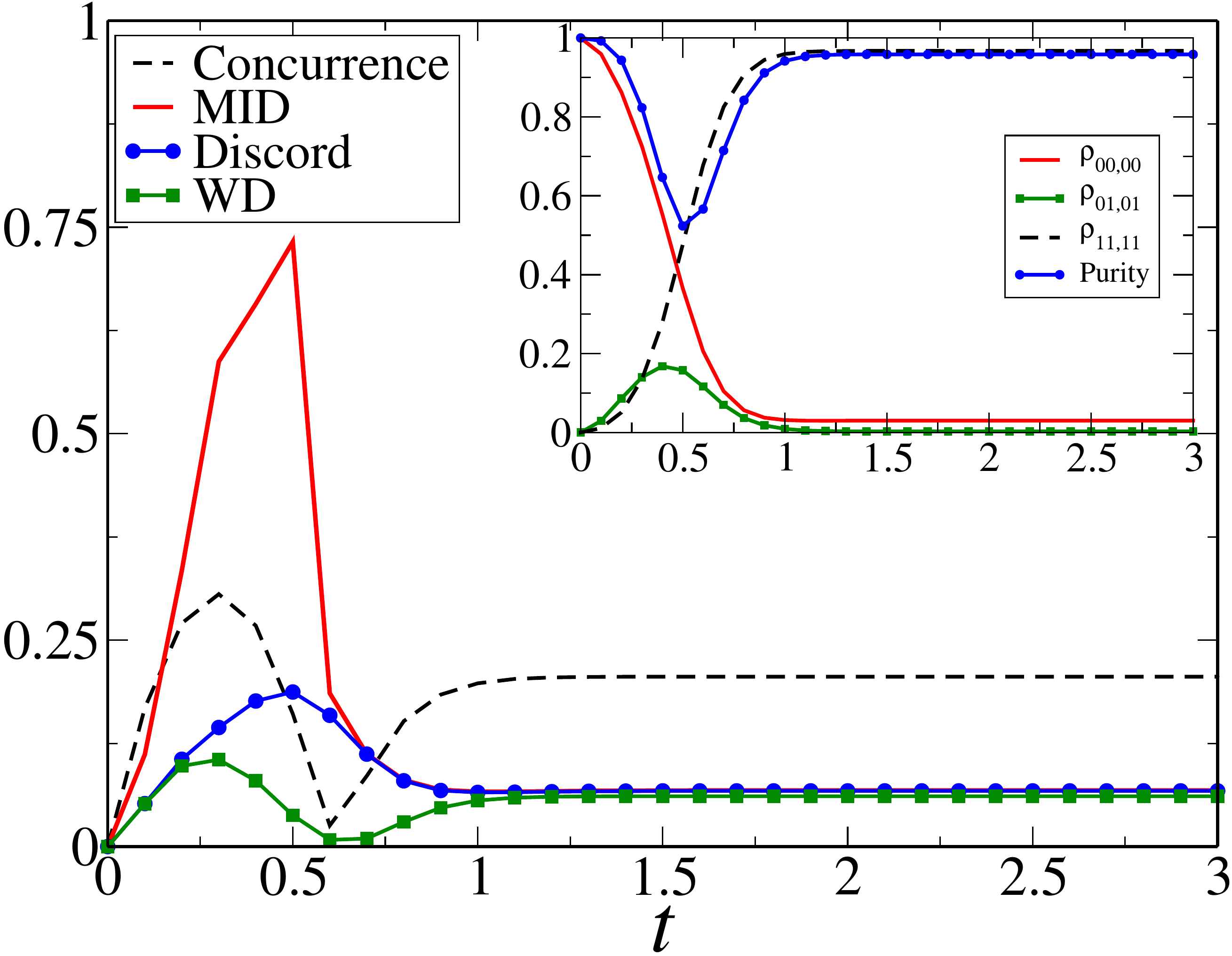}\label{fig5}
\caption{(Color online) The time evolution for MID (red continuous), discord (blue circles), work deficit (green squares) and concurrence (black dashed) for the V configuration. The field detunings are $\Delta_1$ = $\Delta_2$ = 0, and the phases of the Rabi frequencies are $\phi_1$ = $\phi_2$ = 0. The level decay rates are $\Gamma_1 = \Gamma_2$ = 6.0. SPA for this configuration requires that $\Omega_{1,2} < \Gamma_{1,2}$. The driving field strengths are $\Omega_1$ = 2.0, $\Omega_2$ = 4.0. The inset shows the evolution of population elements of the two-photon density matrix and its purity.}
\label{fig5}
\end{center}
\end{figure}

Similar parameter regimes can also be generated in the $\Lambda$ and V configuration as shown in Fig.\,\ref{fig4} and Fig.\,\ref{fig5}, respectively. Interestingly, the monotonic nature of the correlations in the $\Lambda$ and V configuration is different from that of the $\Xi$ configuration. In the relatively high ground state excitation regime (high $\Omega_1$) in Fig.\,\ref{fig4} we observe that the correlations attain steady-state values faster than in the $\Xi$ configuration. The measurement-based correlations such as QD and WD are not monotonic with either concurrence or MID at smaller times unlike in the $\Xi$ configuration where concurrence is always monotonic with QD and WD. There is a temporal discontinuity of concurrence around $t \approx$ 0.5. The concurrence collapses to a small finite value before reviving sharply. The revival of entanglement is associated with an increase in discord in the vicinity of the collapse. Such a feature of the correlations has been reported in other systems \cite{HSD}. Other measures do not exhibit any discontinuity. At greater times ($t \geq$ 1.0), the correlations are steady and weakly monotonic. The behavior of the correlations is again related to the population dynamics and purity of the density matrix as evident from Fig.\,\ref{fig4} and Fig.\,\ref{fig5}.

The different behavior of the correlation monotonicity in the $\Lambda$ and V configuration as compared to $\Xi$ can be understood from the structural difference in the arrangement of the atomic levels. The highest excited level in the $\Xi$ configuration is the metastable state with a decay rate $\Gamma_2\approx$ 1.0. In contrast, the $\Lambda$ and V configurations have $\Gamma_2 \approx$ 6.0. Hence, the evolution of population dynamics and the temporal steady state occurs faster ($t \approx$ 0.5) than in the $\Xi$ configuration. However, the steady state bounds of MID or concurrence in different parameter regimes are common to all the configurations. From Fig.\,\ref{fig5}, we observe that the steady state population dynamics in the V system results in high purity. We get an overlap of all measurement based correlations monotonically bounded by a low concurrence. This is due to the uniform decay rates of the two excited levels leading to a uniform distribution of population probabilities between $|00\rangle$ and $|11\rangle$ with negligible population in $|01\rangle$.

\begin{figure}[ht]
\begin{center}
\includegraphics[scale=0.52]{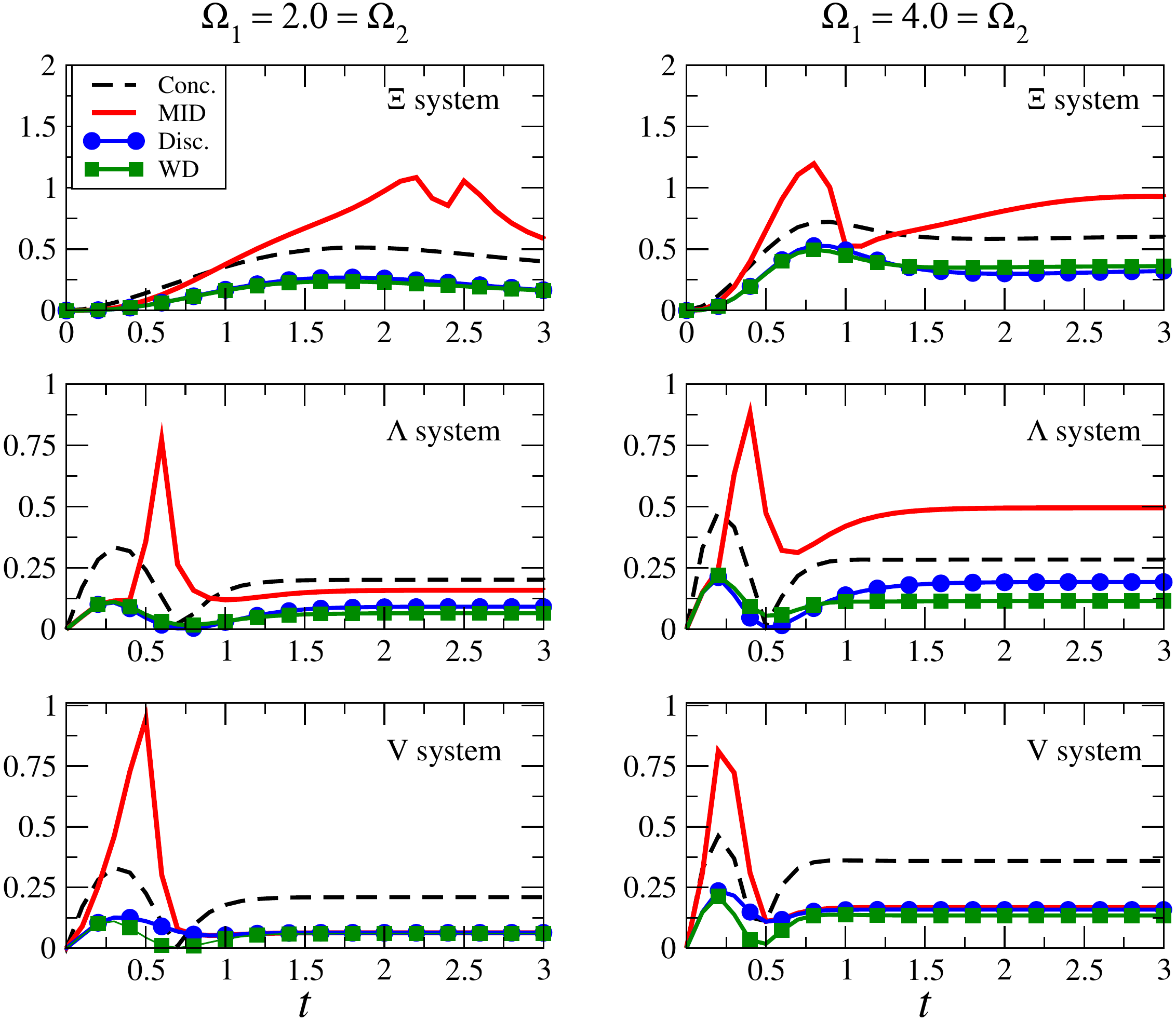}\label{fig6}
\caption{(Color online) Time evolutions of MID (red continuous), discord (blue circles), work deficit (green squares) and concurrence (black dashed) for all the three configurations, $\Xi$ (top), $\Lambda$ (middle) and V (bottom). The field detunings are $\Delta_1$ = $\Delta_2$ = 0, and the phases of the Rabi frequencies are $\phi_1$ = $\phi_2$ = 0. The chosen driving field strengths of $\Omega_1 = \Omega_2$ = 2.0 (left panel) and $\Omega_1 = \Omega_2$ = 4.0 (right panel) satisfy the SPA for all three configurations.}
\label{fig6}
\end{center}
\end{figure}

In Fig.\,\ref{fig6}, we present a comparative study of the behavior in the three configurations under investigation, by choosing a common parameter regime that satisfies the SPA for all configurations. We consider a region of moderately low values of the driving fields ($\Omega_1 = \Omega_2 = 2.0$) and another region of higher values ($\Omega_1 = \Omega_2 = 4.0$). Some of the aspects of the correlation that can be qualitatively studied are monotonicity, temporal steady state, qualitative hierarchy and the nature of the two-photon density matrix.

The correlation dynamics of the $\Xi$ system is dominated by the population dynamics of the photon state which evolves relatively slow due to the metastable nature of the highest excited level. We observe that the correlations do not achieve temporal steadiness in the observed times. This also leads to a lack of genuine monotonicity, with MID forming an upper bound in both the high and low field regimes. In comparison, the $\Lambda$ and V systems have less stable excited states and hence achieve steady state correlations faster. For the $\Lambda$ system, steady state correlations are fairly monotonic with concurrence bound at lower fields and MID bound at higher fields. V systems have relatively low values of steady-state quantum correlations, as compared to the long time values in $\Xi$ systems, with a concurrence bound at all fields.

\section{Summary}
\label{sec6}
The generation of mixed state quantum correlations is an important problem in the context of future implementation of quantum information tasks. The characterization of such correlations in a possible experimentally realizable model is thus an worthwhile exercise. With developments in experimental techniques and measurements of information-theoretic correlations like QD \cite{QDexp}, the importance of generating and analyzing correlations has manifestedly increased in recent times. The possibility of using well studied quantum optical systems like the three-level atom to generate, characterize and parametrically control mixed state quantum correlations is indeed an encouraging step in this direction.

In this paper, we have used a semiclassical three-level atom interacting with classical driving fields to generate a correlated two-mode photon pair controlled by the driving field parameters and the phenomenological decay terms of the atomic level. Under certain physical approximations, we have exhaustively studied and compared the dynamical behavior of various nonclassical correlations of the two-photon state generated in the semiclassical atom+field system. We have probed the dynamics of entanglement (concurrence) as well as of measurement-based correlations, such as MID, QD and WD, for three different configurations, namely, $\Xi$, $\Lambda$ and V, of the three-level atom driven by two controlled external classical driving fields. The qualitative characterization of the correlations based on the monotonicity, general hierarchy and steady state behavior are achieved using the field parameters and the decay terms.

The correlation behavior is observed to be configuration dependent. The manipulation of control parameters in different configurations leads to variations in the dynamic evolution of the correlations. The $\Xi$ configuration produces photon states with relatively high correlation even at low driving fields. The dynamics in the $\Xi$ system is more mixed as compared to the $\Lambda$ and V systems. The $\Xi$ system can be suitably controlled using the driving fields to generate correlations dominated both by MID or concurrence and is ideally suited to experimentally study the temporal evolution of the two measures with respect to the evolution of the system in the Hilbert space. The optimization of MID and monotonicity of measurement based correlations can be experimentally analyzed using quantum state tomography. $\Lambda$ and V systems are better suited for generating steady monotonic correlations in both low and high strength driving field regimes. The $\Lambda$ system can be suitably tuned to generate steady correlations with either entanglement or MID as an upper bound. The mixedness in the generated states can be controlled using the driving field strengths for implementation in experiments. V systems, however, can generate ideally pure correlated photons bounded by concurrence at all field strength regimes. The absence of a metastable state in the V system allows production of pure correlated output photon states. The measurement-based correlations are all equal at steady values. However, in the $\Lambda$ and V systems, significant correlation is generated only for high driving field strengths.

Hence, specific regimes and configurations can be used to generate and manipulate the correlations in the output two-photon state as desired. Our findings may thus be immensely useful in practical implementations with such interacting photon states.

\section*{Acknowledgments}
The work of HSD is supported by the University Grants Commission, India. HSD also thanks the Indian Institute of Technology Rajasthan for hospitality during a visit. AC thanks the National Board of Higher Mathematics, Department of Atomic Energy, Government of India, for financial support.


\begin{thebibliography}{99}
\bibitem{comm1} The atomic density matrix is a 3$\times$3 matrix pertaining to the three-dimensional single atom. The two-mode photon density matrix is a 4$\times$4 matrix, since each photon has two orthogonal states $|0\rangle$ and $|1\rangle$. The equivalence is possible because the atom-field interaction ensures that the photon density matrix has a Schmidt decomposition that reduces its rank to 3.
\bibitem{comm2} It is observed that the qualitative nature of the correlation dynamics is not very sensitive to finite detunings.
\end{thebibliography}
\end{document}